\documentclass[pra,twocolumn,superscriptaddress,nofootinbib]{revtex4-1}
\usepackage{dcolumn}
\usepackage{amssymb}
\usepackage{amsmath}
\usepackage{siunitx}
\usepackage{soul}
\usepackage[colorinlistoftodos,prependcaption,textsize=tiny]{todonotes}
\usepackage{blindtext}
\usepackage{multirow}
\usepackage{float}
\presetkeys%
    {todonotes}%
    {inline,backgroundcolor=yellow}{}

\begin{document}
\title{Equilibration of the planar modes of ultracold two dimensional ion crystals in a Penning trap}
\author{Chen Tang}
\affiliation{Department of Physics, University of Colorado at Boulder}
\author{Athreya Shankar}
\affiliation{Center for Quantum Physics, Faculty of Mathematics, Computer Science and Physics, University of Innsbruck, Innsbruck A-6020, Austria}
\affiliation{Institute for Quantum Optics and Quantum Information, Austrian Academy of Sciences, Innsbruck A-6020, Austria.}
\author{Dominic Meiser}
\affiliation{Department of Physics, University of Colorado at Boulder}
\author{Daniel H. E. Dubin}
\affiliation{Department of Physics, University of California San Diego}
\author{John J. Bollinger} 
\affiliation{Time and Frequency Division, National Institute of Standards and Technology}
\author{Scott E. Parker}
\affiliation{Department of Physics, University of Colorado at Boulder}

\begin{abstract}

Planar thermal equilibration is studied using direct numerical simulations of ultracold two-dimensional (2D) ion crystals in a Penning trap with a rotating wall.  The large magnetic field of the trap splits the modes that describe in-plane motion of the ions into two branches: High frequency cyclotron modes dominated by kinetic energy and low frequency $\mathbf{E \times B}$ modes dominated by potential energy associated with thermal position displacements. Using an eigenmode analysis we extract the equilibration rate between these two branches as a function of the ratio of the frequencies that characterize the two branches and observe this equilibration rate to be exponentially suppressed as the ratio increases.  Under experimental conditions relevant for current work at NIST, the predicted equilibration time is orders of magnitude longer than any relevant experimental timescales. We also study the coupling rate dependence on the thermal temperature and the number of ions. Besides, we show how increasing the rotating wall strength improves crystal stability. These details of in-plane mode dynamics help set the stage for developing strategies to efficiently cool the in-plane modes and improve the performance of single-plane ion crystals for quantum information processing.

\end{abstract}
\maketitle
\section{Introduction}

Single-plane crystals of several hundred ions in Penning traps provide an appealing platform for quantum information processing and quantum sensing. The large number of qubits in this system provides for the possibility of quantum simulations of paradigmatic spin and spin-boson models in a regime where classical simulation becomes intractable \cite{britton2012engineered,safavi2018prl,wang2013,cohn2018njp}. Experimental work to date has focused on all-to-all interactions between the ion qubits, studying the buildup of qubit correlations in a regime where experiment can be benchmarked with theory \cite{bohnet2016quantum,garttner2017measuring}, but with improved control and the addition of techniques such as single-site addressability, more complex simulations and general information processing will be possible~\cite{Schindler_2013}.  This promise has motivated recent efforts to improve the Penning trap platform and increase the control and tools available to the experimentalist.  This includes efforts to develop miniaturized permanent-magnet systems that offer portability \cite{mcmahon2020PRA}, traps with improved optical access \cite{ball2019RSI}, the incorporation of sideband cooling \cite{hrmo2019PRA}, and proposals for quantum computing and simulation in arrays of Penning traps \cite{jain2020scalable}.  

In trapped-ion quantum information processing, strong interactions between the ion qubits (or spins) are generated by coupling the ion crystal spin degrees of freedom with the ion crystal motional (or mode) degrees of freedom through the application of a spin-dependent force.  For single-plane crystals in Penning traps, this is routinely accomplished by coupling the ion spins to the drumhead modes that describe ion motion perpendicular to the plane of the crystal (or parallel to the magnetic field of the Penning trap) \cite{sawyer2012PRL}.  A single-plane crystal with $N$ ions will support $N$ drumhead modes, each of which can be described as a simple harmonic oscillator. The drumhead modes are efficiently cooled to near their ground state by a combination of Doppler and EIT (electromagnetically induced transparency) cooling \cite{jordan2019PRL,AthreyaEIT}.

In contrast, the in-plane ion motion is complicated by the presence of the strong magnetic field of the trap and has not to date been employed for quantum information processing tasks. The strong magnetic field splits the planar normal modes into a cyclotron branch containing $N$ high frequency modes and an $\mathbf{E \times B}$ branch containing $N$ low frequency modes.  Additionally, the planar modes do not undergo simple harmonic motion and their average potential and kinetic energies are not equal. The $\mathbf{E \times B}$ modes are dominated by potential energy associated with thermal position displacements, while the cyclotron modes are dominated by kinetic energy associated with cyclotron motion. In contrast to the drumhead modes, efficient cooling of the in-plane modes has not been demonstrated experimentally or even clearly discussed theoretically for multi-ion crystals. Doppler cooling of the cyclotron modes to millikelvin temperatures appears feasible \cite{tang2019,torrisi2016PRA}, but recent theoretical work indicates that observed frequency instabilities of the drumhead mode spectrum can be attributed to an elevated temperature of order $\SI{10}{\milli\kelvin}$ for the $\mathbf{E \times B}$ modes \cite{athreya2020}. A detailed understanding of the planar mode dynamics and the energy exchange between the different planar mode branches, besides being of fundamental importance~\cite{Glinsky_1992, Jensen2005, Dubin2005PRL, AndereggPRL2009}, is an important first step in the design of efficient cooling techniques as well as quantum information protocols that utilize these modes.

In this paper, we investigate the exchange of energy between the cyclotron and $\mathbf{E \times B}$ branches of single-plane ion crystals in Penning traps using an eigenmode analysis of a first-principle molecular dynamics-type simulation \cite{tang2019}.  We characterize the energy exchange as a function of the ratio $R$ of the ion crystal cyclotron and $\mathbf{E \times B}$ center-of-mass mode frequencies (see Eq. (\ref{equ:ratio})).  The center-of-mass frequencies provide a convenient characterization for the frequency ratio between the two branches. From simulations performed with $5<R<10$ we find that the exchange of energy between the two branches is exponentially suppressed as a function of $R$.  A simplistic extrapolation to $R\approx735$, relevant for the current NIST experimental set-up, gives an equilibration time many orders of magnitude longer than the age of the universe. In addition, we also study the less-sensitive dependence of the rate of energy exchange between the branches on the initial energy and the number of trapped ions. 

Finally, for large $R$ where the energy exchange between branches is negligible, we study the exchange of energy between modes within a given branch and observe a significantly faster equilibration within the $\mathbf{E \times B}$ branch than the cyclotron branch. In the course of the above studies, we also show that increasing the rotating wall strength leads to improved crystal stability. These observations improve our understanding of the in-plane mode dynamics, setting the stage for developing strategies for efficiently cooling the $\mathbf{E \times B}$ modes.  The isolation of the cyclotron modes suggests their potential use and efficacy in quantum information processing protocols.

The organization of the paper is as follows. In Sec. \ref{sec:theory}, we review the governing equations for the rotating-wall Penning trap configuration at NIST. The model equations are the starting point for both direct numerical simulation and the linear eigenmode analysis.  In Sec. \ref{sec:energy}, we present both an eigenmode and band-pass filter technique for determining the energies of the two mode branches. The eigenmode technique is based on linearizing the system, details of which are presented in Appendix \ref{app:eigenmode}. In Sec. \ref{sec:coupling} we discuss Penning trap and ion crystal parameters that affect the coupling rate and develop a systematic procedure for obtaining different crystal configurations characterized by the desired parameters. In Sec. \ref{sec:stability}, we study the influence of the rotating wall strength on the ion crystal stability. We find that a strong rotating wall improves the crystal stability and the effectiveness of the eigenmode measurement. In Sec. \ref{sec:simulation}, we present the first-principles simulation results. We begin by showing a thermalization process of the modes for $R=5$, where equipartition of the mode energies is reached after 10 ms evolution. We then study the dependence of the equilibration rate between the cyclotron and $\mathbf{E \times B}$ modes on several parameters in Sec. \ref{subsec:coupling_rate}. For large $R$, where the inter-branch coupling is very weak, we also examine coupling among the modes within each branch. Finally, in Sec. \ref{sec:discussion}, we summarize with a discussion and concluding remarks.

\section{\label{sec:theory}Theoretical Formulation}

We have developed an $N$-particle classical simulation of ultra-cold ions in a Penning trap, including a rotating wall and axial and planar Doppler cooling \cite{tang2019}. The code includes a fairly realistic implementation of the experimental configuration employed at NIST \cite{britton2012engineered,bohnet2016quantum,sawyer2014spin}. Here we use this code (without implementing the laser cooling) to simulate the equilibration of the planar modes. We analyze the simulation through an eigenmode decomposition. In this section, we introduce the model and parameters relevant for single-plane crystals in Penning traps and describe the planar normal modes of motion \cite{dan2020,athreya2020}. Details of the normal mode analysis are given in Appendix \ref{app:eigenmode}.

We treat $N$ ions, all with the same mass $m$ and charge $q$, as classical point particles confined in a rotating-wall Penning trap. The Penning trap confining fields consist of a magnetic field $\mathbf{B}=B\hat{z}$, a quadrupole electrostatic potential $\varphi_{trap}(\mathbf{x})=\frac{1}{4}k_z(2z^2-x^2-y^2)$, and a time-dependent potential $\varphi_{wall}(\mathbf{x},t)=\frac{1}{2}k_z\delta\left( x^2 - y^2\right)\cos\left [2\left(\theta +\omega_R t\right)\right]$ called the rotating wall. The dimensionless parameter $\delta$ characterizes the relative strength of the rotating wall potential to that of $\varphi_{trap}$. The parameters $\theta$ and $\omega_R$ are the azimuthal angle and the rotating wall frequency. Further details of the simulation model are given in Ref.~\cite{tang2019}. 

Experimentally~\cite{britton2012engineered,bohnet2016quantum,sawyer2014spin}, the ions are cooled to a regime where the ions are strongly correlated with a correlation coefficient $\Gamma = q^2/(4\pi\epsilon_0 ak_BT_p) \gg 1$~\cite{dan1993} ($a$ is the typical inter-ion spacing and $T_p$ is the temperature). The strongly correlated ions form a crystal that rigidly rotates at the frequency $\omega_r=\omega_R$~\cite{tang2019} as the rotating wall potential locks the ion crystal rotation frequency. In the rotating frame of the crystal, the potential energy of $N$ ions with coordinates $\mathbf{x}_{i}=(x_i,y_i,z_i)$ is time independent and is given by~\cite{tang2019}
\begin{equation}
\begin{aligned}
\Psi_r=&\sum_{i=1}^N \frac{1}{2}m \omega_z z_i^2 \\
&+\sum_{i=1}^N \frac{1}{2} m\left[\omega_r(\Omega-\omega_r)-(1/2)\omega_z^2\right] (x_i^2+y_i^2) \\
&+\sum_{i=1}^N \frac{1}{2}m\delta\omega_z^2 (x_i^2- y_i^2)\\ 
&+\sum_{i=1}^N  \sum_{j \ne i}\frac{q^2}{8\pi\varepsilon_0}\frac{1}{\left| \mathbf{x}_{i} - \mathbf{x}_{j} \right|},
\end{aligned}
\label{equ:HEP}
\end{equation}
where we parametrize the trap strength with axial trapping frequency $\omega_z=\sqrt{qk_z/m}$ and bare cyclotron frequency $\Omega=qB/m$. A stationary equilibrium ion crystal state with positions $\mathbf{x}_{0i}$ satisfying $\partial \Psi_r/\partial \mathbf{x}_{0i}=0$ for $(i=1,2,...,N)$ can be found numerically by minimizing the potential energy $\Psi_r$ \cite{wang2013}. 

In this work, we study single-plane ion crystals that have a two-dimensional structure. There are two requirements on the radial confinement strength for an ion crystal to maintain a single-plane structure in a Penning trap. The strength of the radial confinement (second line in Eq. (\ref{equ:HEP})) relative to the strength of the axial confinement (first line in Eq. (\ref{equ:HEP})) is characterized by the parameter
\begin{equation}
  \beta=\omega_r(\Omega-\omega_r)/\omega_z^2-1/2.
\end{equation}
For a single plane crystal of $N$ ions, the trap asymmetry $\beta$ must be less than a critical $\beta_c(N)$ \cite{dan1993},
\begin{equation}
    \beta < \beta_c(N) \approx 0.665/\sqrt{N} .
\end{equation}
Second, the radial confinement strength must be stronger than the rotating wall strength.  The force along the $y$ axis is 
\begin{equation} 
F_y= - (\beta-\delta)m\omega_z^2y,
\end{equation}
which requires $\beta>\delta$ for trapping along the $y$ direction.

At ultracold temperatures, ion displacements relative to the equilibrium inter-ion spacing are small. This feature allows us to linearize the ion motion and then solve for the normal modes. In a two-dimensional crystal, the linearized ion motion in the out-of-plane ($z$) direction decouples  from that in the planar ($x$ and $y$) direction. In this work, we solve for the normal modes in the planar direction. As presented in Appendix \ref{app:eigenmode}, there are $2N$ normal modes in the planar direction with eigenvectors $\mathbf{u}_n$ and mode frequencies $\omega_n$. In terms of these eigenvectors, we can express any small position and velocity displacements $\mathbf{s}_\perp=(\mathbf{x}_{\perp} ,\mathbf{v}_{\perp})^T$ of the ions in the planar direction as
\begin{equation}
\begin{aligned}
\mathbf{s}_\perp=\sum_{n=1}^{2N}a_n e^{-i\omega_n t}\mathbf{u}_n + \sum_{n=1}^{2N}a_n^* e^{i\omega_n t}\mathbf{u}_n^*.
\end{aligned}
\label{equ:decomposition}
\end{equation}

Because of the Lorentz force arising from the magnetic field, the planar eigenvectors obey a generalized orthogonality relation with respect to a composite energy matrix $\mathbb{E}$~\cite{athreya2020}. As shown in Eq.~(\ref{equ:E_matrix}), $\mathbb{E}$ is constructed out of the (diagonal) mass matrix of the ions and a stiffness matrix $\mathbb{K}_\perp$ that is obtained by linearizing the ion equations of motion. As a result of the $\mathbb{E}$-orthogonality of the eigenvectors, the complex amplitude $a_n$ of each normal mode is given by  
\begin{equation}
a_n= \mathbf{u}_n^*\mathbb{E}\mathbf{s}_\perp,
\label{equ:amp}
\end{equation}
where the eigenvectors $\mathbf{u}_n$ have been normalized according to Eq.~(\ref{eqn:un_norm}).

\begin{figure}
\includegraphics[width=9cm]{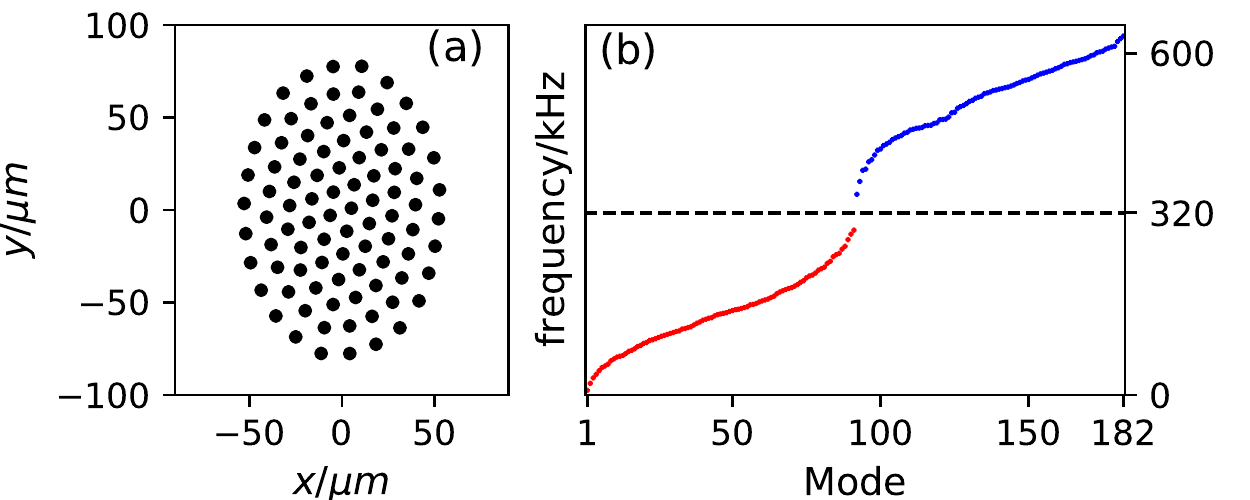}
 \caption{(a) Configuration of a crystal with $N=91$ ions. Relevant trap and ion parameters are $\omega_z=2\pi\times \SI{704}{\kilo\hertz}$, $\omega_r= 2\pi\times \SI{400}{\kilo\hertz}$, $\beta=$ 0.05, $\delta=$ 0.0126, $B=\SI{4.4588}{\tesla}$, $m=$ 63.3 u, and $q=e$. These parameters give rise to $R=5$ (defined in Eq. \ref{equ:ratio}). (b)  The $2N = 182$ planar mode frequencies of the crystal in the $\mathbf{E \times B}$ (red) and cyclotron (blue) branches. The dashed line shown is at $320$ kHz.}
 \label{fig:Eval}
\end{figure}

Among the $2N$ normal modes, $N$ modes correspond to the low-frequency $\mathbf{E \times B}$ branch and $N$ modes correspond to the high-frequency cyclotron branch (which we will respectively denote by subscripts $b$ and $c$ in what follows).  We arrange the $2N$ modes ($n$ from $1$ to $2N$) in ascending order according to their frequencies. The $\mathbf{E \times B}$ branch then contains modes $1$ to $N$ and the cyclotron branch contains modes $N+1$ to $2N$. A typical configuration of an ion crystal studied in this manuscript is shown in Fig.~\ref{fig:Eval}(a). The associated distribution of the mode frequencies is presented in Fig.~\ref{fig:Eval}(b). Here trap and ion parameters are chosen so that the frequencies of the two branches are separated by a small amount. 

\section{\label{sec:energy}Diagnostic tools for in-plane modes}

In this section, we first describe the eigenmode analysis method that we use to measure the kinetic and potential energies of individual planar modes in the course of a molecular dynamics simulation. Under certain conditions, the linearization assumption giving rise to the mode picture can be marginal due to the significant potential energy (and displacements of the ions) associated with the $\mathbf{E \times B}$ modes. Therefore, we subsequently also discuss a band-pass filter method, based on the Fourier transform of a time-series of the ions' velocities, which is applicable regardless of the linearization assumption. We use the latter method to validate the results from our eigenmode analysis.

\subsection{Eigenmode measurement method}
We start by separating eigenvectors into their coordinate and velocity components as $\mathbf{u}_n=(\mathbf{r}_n, \mathbf{v}_n)^T$.  Using Eqs.~(\ref{equ:E_planar}) and (\ref{equ:decomposition}), we express the total in-plane thermal energy in terms of the planar modes as 
\begin{equation}
\begin{aligned}
E&=
\sum_{n=1}^{2N} \left| a_n \right|^2 (\mathbf{r}_n^* \mathbb{K}_{\perp}\mathbf{r}_n + m\mathbf{v}_n^* \mathbf{v}_n) \\
&=\sum_{n=1}^{2N}k_BT_n\equiv  2Nk_BT_p.
\end{aligned}
\label{equ:EPK}
\end{equation}
Here, $T_p$ is defined as a mean planar temperature while $T_n$ describes the temperature of a single mode. In Eq. (\ref{equ:EPK}), terms involving $\mathbf{r}_n$ and $\mathbf{v}_n$ respectively represent the potential ($E_p^n$) and kinetic ($E_k^n$) energies in a single mode. We replace $a_n$ by Eq. (\ref{equ:amp}) to obtain the potential and kinetic energies in a single mode as 
\begin{equation}
\begin{aligned}
E_p^n &= \left |\mathbf{u}_n^*\mathbb{E}\mathbf{s}_\perp \right |^2\mathbf{r}_n^* \mathbb{K}_{\perp}\mathbf{r}_n,\\
E_k^n &= \left |\mathbf{u}_n^*\mathbb{E}\mathbf{s}_\perp \right |^2 m\mathbf{v}_n^* \mathbf{v}_n.\\
\end{aligned}
\label{equ:EM_measure}
\end{equation}
Equation (\ref{equ:EM_measure}) allows measurement of the mode potential and kinetic energies of any instantaneous state $\mathbf{s}_\perp(t)$ based on the orthonormal eigenvectors set $\{\mathbf{u}_n\}$. To evaluate the energy distribution during an evolution process, we simulate the crystal evolution and record ion displacements $\mathbf{s}_i(n_s\Delta t)=(\mathbf{x}_i(n_s\Delta t),\mathbf{v}_i(n_s\Delta t))$ in the rotating frame with the sampling period $\Delta t$ and total sample number $N_s$. Using the recorded velocities and displacements, we calculate the kinetic energies in the two branches based on Eq. (\ref{equ:EM_measure}) as 
\begin{equation}
\begin{aligned}
K_b(n_s \Delta t) =& \sum_{n\in b}\left |\mathbf{u}_n^*\mathbb{E}\mathbf{s}_i(n_s \Delta t) \right |^2 m\mathbf{v}_n^* \mathbf{v}_n,\\
K_c(n_s \Delta t) =& \sum_{n\in c}\left |\mathbf{u}_n^*\mathbb{E}\mathbf{s}_i(n_s \Delta t) \right |^2 m\mathbf{v}_n^* \mathbf{v}_n.
\end{aligned}
\label{equ:Ek_EIG}
\end{equation}
Similar expressions are obtained for the potential energies in the two branches by replacing $m\mathbf{v}_n^* \mathbf{v}_n$ with $\mathbf{r}_n^* \mathbb{K}_{\perp}\mathbf{r}_n$ in Eq. (\ref{equ:Ek_EIG}).

\subsection{Band-pass filter method}

A second method for measuring the kinetic energies in the two mode branches is by band-pass filtering the velocities as described below. For the same recorded velocities used in Eq. (\ref{equ:Ek_EIG}), we perform a Fourier transform on the velocity of ion $j$ by means of 
\begin{equation}
\begin{aligned}
  \tilde{\mathbf{v}}_j(\omega) = \frac{1}{\tau}\int_0^\tau e^{-i\omega t}\mathbf{v}_j(t)\, \hbox{d}t,
\end{aligned}
\end{equation}
where $\tau=N_s\times\Delta t$ is the total recording time. Given the discretely sampled velocities, we approximate the Fourier transform by utilizing a discrete Fast Fourier Transform
\begin{equation}
\begin{aligned}
 &\tilde{\mathbf{v}}_j(l\Delta\omega) = \frac{1}{\tau}\int_0^\tau e^{-il\Delta\omega t}\mathbf{v}_j(t)\,\hbox{d}t \\
    &\approx \frac{1}{N_s}\sum_{n_s=1}^{N_s}e^{-i2\pi ln_s/N_s}\mathbf{v}_j(n_s \Delta t),
\end{aligned}
\label{equ:FFT}
\end{equation}
where $l\in\{0,1,...,N_s-1\}$ and $\Delta \omega=2\pi/\tau$ is the frequency resolution. In order to accommodate the full frequency range of the planar modes, $N_s \Delta\omega/2=\pi/\Delta t$ exceeds the maximum mode frequency $\omega_m$. 

We then apply a band-pass filter to separate $\tilde{\mathbf{v}}_j(l\Delta\omega) $ with respect to mode frequency. We choose the band-pass filter frequency $l_0 \Delta \omega$, with $l_0$ a positive integer, to be located in the frequency gap of the two mode branches, i.e. 
\begin{equation}
\mathrm{max}\{\omega_b\}< l_0 \Delta\omega <\mathrm{min}\{\omega_c\}.\end{equation}
With the help of $l_0$, we divide $ \tilde{\mathbf{v}}_j(l\Delta\omega)$ into
\begin{equation}
\begin{aligned}
  \tilde{\mathbf{v}}_j(l\Delta\omega) \rightarrow  \tilde{\mathbf{v}}_j^b(l<l_0) + \tilde{\mathbf{v}}_j^c (l\geq l_0).
\end{aligned}
\end{equation}
Next, we apply inverse Fourier transforms to transform $\tilde{\mathbf{v}}_j^b$ and $\tilde{\mathbf{v}}_j^c$ back to $\bar{\mathbf{v}}_j^b(n_s \Delta t)$ and $\bar{\mathbf{v}}_j^c(n \Delta t)$ in the time domain. We repeat the above process for all ions ($j=1,...,N$) to calculate the kinetic energies, $K_b(n_s \Delta t)$ and $K_c(n_s \Delta t)$, in the two branches as
\begin{equation}
\begin{aligned}
K_b(n_s \Delta t) =& \frac{m}{2}\sum_{j=1}^{N}\left|\bar{\mathbf{v}}_j^b(n_s \Delta t)\right|^2,\\
K_c(n_s \Delta t) =& \frac{m}{2}\sum_{j=1}^{N}\left|\bar{\mathbf{v}}_j^c(n_s \Delta t)\right|^2.
\end{aligned}
\label{equ:Ek_FFT}
\end{equation}
We can then compare the results from Eqs. (\ref{equ:Ek_FFT}) and (\ref{equ:Ek_EIG}) in the simulation for validation purposes. In Sec. \ref{subsec:benchmarking}, good agreement between the two methods is achieved when the displacements are small and no slippage or distortion of the crystal is observed. We have also found good agreement between the total kinetic and potential thermal energies obtained from the eigenmode analysis and those obtained from a direct evaluation using the ion coordinates in the simulation, again for small displacements. 

It is worth noting that the eigenmodes method is not restricted by the requirement of the sampling period and the size of the data collection. The band-pass filter method, however, requires an appropriate sampling period and enough data to cover the frequencies of all planar modes. While the band-pass filter method only measures the kinetic energy, it performs better than the eigenmode measurement when displacements are not extremely small.

\section{\label{sec:coupling}Parameters controlling equilibration}

In this section, we identify important trap and crystal parameters that control the thermal equilibration of the planar modes. We also describe a procedure to tune the parameters and obtain similar crystal configurations whose equilibration rates can be meaningfully compared.

Normal modes of trapped ion crystals are only decoupled in the limit of small-amplitude displacements. In reality, anharmonic terms in the Coulomb interaction couple different modes and may eventually lead to equilibration~\cite{schiffer1991way}. Prior work with one-dimensional ion chains in an RF Paul trap showed that the equilibration rate between the high-frequency radial modes and the low-frequency axial modes is exponentially suppressed in the ratio of the characteristic frequencies of motion along these two directions~\cite{shijie1993}. This result can be understood via energy conservation in a phonon picture, wherein for a large separation of frequencies, several low-frequency phonons must be created in order to annihilate a single high-frequency phonon. Such multiple phonon processes arise as high-order terms in the Coulomb interaction with small effective rates.

A natural measure of the characteristic frequency of motion for the cyclotron and $\mathbf{E \times B}$ branches is provided by the center-of-mass (c.m.) frequencies $\omega_+, \:\omega_-$ of each branch.  The c.m. frequencies are independent of ion number and are the same as the single-ion motional frequencies. In the weak rotating wall limit ($\delta \ll 1$), we can solve analytically for the two frequencies to obtain, in a frame rotating at a frequency $\omega_r$, 
\begin{equation}
\omega_\pm = \frac{\sqrt{\Omega^2-2\omega_z^2}\pm (\Omega-2\omega_r)}{2}.
\end{equation}
Here $\omega_+$ is the c.m. mode for the cyclotron branch and $\omega_-$ is the center-of-mass mode for the $E \times B$ branch. We study the dependence of the equilibration between the two branches on the ratio
\begin{equation}
    R = \frac{\omega_+}{\omega_-}.
\label{equ:ratio}
\end{equation}

Other important parameters that can impact the equilibration rate are the number $N$ of trapped ions and the thermal temperature in the planar direction. With larger numbers of ions, one expects more available modes for satisfying the frequency match required for phonon-phonon coupling. When the temperature is higher, the ion displacements are larger and anharmonic Coulomb coupling is stronger~\cite{hasse1992relaxation}. 

\begin{figure}
\includegraphics[width=8.5cm]{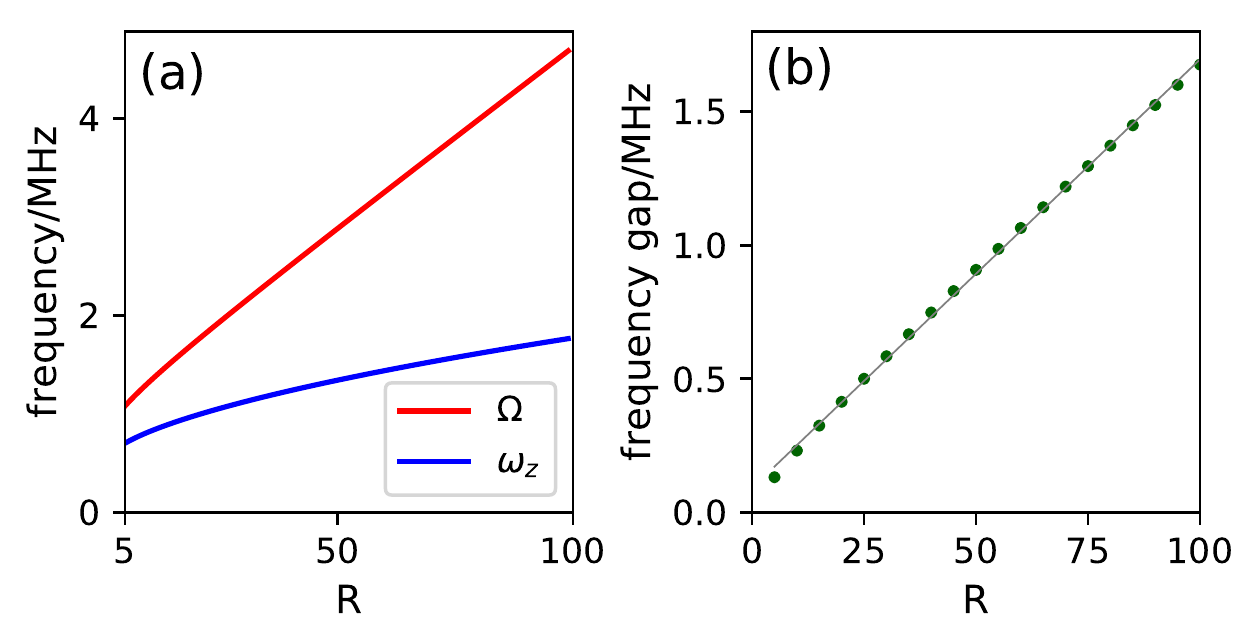}
 \caption{(a) The trap axial frequency $\omega_z$ and ion cyclotron frequency $\Omega$ as functions of $R$ with $B=4.4588$ T, $\omega_r=2\pi\times 400$ kHz, $\beta=0.05$, and $\delta=0.0126$ held fixed. (b) Frequency gap $\Delta_\omega$ as a function of $R$. Other parameters are the same as in Fig. \ref{fig:Eval}. The gray line is a linear fit to the numerical results. }
 \label{fig:para_relation}
\end{figure}

To enable a study of the energy transfer between the planar modes, we develop a systematic procedure by which we can obtain similar crystal configurations that can be meaningfully compared while varying the frequency ratio parameter $R$. This is not trivial due to the large number of trap parameters. We obtain crystals with the same rotation frequency $\omega_r$, magnetic field $B$, the relative rotating wall strength $\delta$ and relative radial confinement strength $\beta$. We study single-plane crystals with $N\leq 127$. The critical trap asymmetry parameter for $N=127$ ions is $\beta_M(127)\approx 0.059$. We fix $\beta=0.05$, $\delta=0.0126$, and $\omega_r=2\pi\times 400$ kHz. The axial trapping frequency $\omega_z$ and the bare cyclotron frequency $\Omega$ can be expressed as functions of $R$, $\beta$, and $\omega_r$ in the following way

 \begin{equation}
\begin{aligned}
\omega_z&=\sqrt{\frac{2}{F(R)}}\frac{\omega_r}{2\beta+1}\left[\sqrt{2\beta+F(R)}+\sqrt{2\beta(1-F(R))}\right],\\
\Omega &= \sqrt{2}\omega_z\sqrt{\frac{2\beta+F(R)}{F(R)}},
\end{aligned}
\end{equation}
where $F(R) = 1-\left(\frac{R-1}{R+1}\right)^2$. The relation between $\omega_z$, $\Omega$, and $R$, with  $\omega_r/(2\pi)=400$ kHz and $\beta=0.05$, is plotted in Fig.~\ref{fig:para_relation}(a). By fixing the rotation frequency and magnetic field we obtain crystals that have approximately the same ion density.  Physically, the cyclotron frequency determines the ion mass through $m=qB/\Omega$ and the axial frequency the required trap voltage for that ion mass. In Fig. \ref{fig:para_relation}(b) we investigate the dependence of the frequency gap between the two branches, $\Delta_\omega=\omega_{\{c, min\}}-\omega_{\{b, max\}}$, on $R$ for $N=91$ ions. The nearly linear relation indicates that $R$ also provides a means of parameterizing the gap between the two branches.

\section{\label{sec:stability}Rotating wall strength}

In order to apply the eigenmode measurement method, we need a stable crystal equilibrium. In this section, we show that a strong rotating wall is a way to achieve such a crystal configuration.
\begin{figure}[h]
\includegraphics[width=\columnwidth]{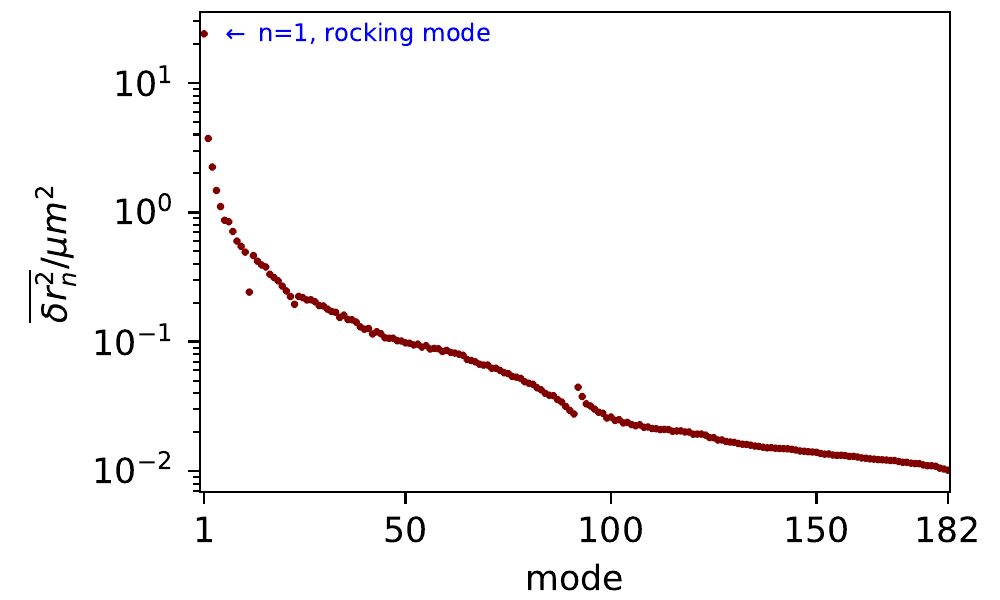}
 \caption{ \label{fig:mean_displacement} Mean squared displacement $\overline{\delta r^2_n}$ of $2N=182$ planar modes with mode temperature $T_n = \SI{1}{\milli\kelvin}$. Here the trap and ion parameters are the same as used in Fig.~\ref{fig:Eval} and discussed in Sec. \ref{subsec:benchmarking}.  In particular $R=5$ and $\delta=0.0126$.}
\end{figure}

As ions with significant displacements escape from the vicinity of their equilibrium positions, the crystal is not stable anymore and we cannot use the eigenmodes associated with the original equilibrium state to describe the system. The failure of the eigenmode method in such situations can be observed numerically from the lack of energy conservation when mode energies are computed using this method. Therefore, the effectiveness of the eigenmode measurement method relies on the crystal stability. To quantify the crystal stability, we consider the sum of the squared thermal displacements in the planar direction of all the ions in the rotating frame. This quantity can be written as a sum of mean-squared thermal displacements $\overline{\delta r^2_n}$ of the individual planar modes, which are given by~\cite{athreya2020}
\begin{equation}
\overline{\delta r^2_n}=\frac{2k_BT_n}{(1+\tilde{R}_n)m\omega_n^2}.
\label{equ:mean_displacement}
\end{equation}
Here $\tilde{R}_n=E_p^n/E_k^n$ is the ratio of the potential to kinetic energy of the $n$th mode,\footnote{We note that $R$ defined in Eq. (\ref{equ:ratio}) can be shown to be the ratio of potential to kinetic energy for the $\mathbf{E \times B}$ c.m. mode \cite{athreya2020}.} $\omega_n$ is the mode frequency, $T_n$ is the mode temperature, and $\delta r^2_n$ is obtained by summing the thermal fluctuations in mode $n$ over all the ions. In Fig.~\ref{fig:mean_displacement}, we plot the distribution of $\overline{\delta r^2_n}$ for $N=91$ and $T_n=\SI{1}{\milli\kelvin}$.  We observe that $\overline{\delta r^2_1}$ of the rocking mode (with $n=1$) is much larger than for the other modes. When the rocking mode temperature gets higher, $(\overline{\delta r^2_1}/N)^{1/2}$ becomes comparable to the inter-particle spacing of $d=\SI{12.1}{\micro\meter}$. Since the contribution to the total crystal displacement is dominated by the rocking mode, we use the mean squared displacement  $\overline{\delta r^2_1}$  of this mode to characterize the crystal stability. 

\begin{figure}[h]
\includegraphics[width=\columnwidth]{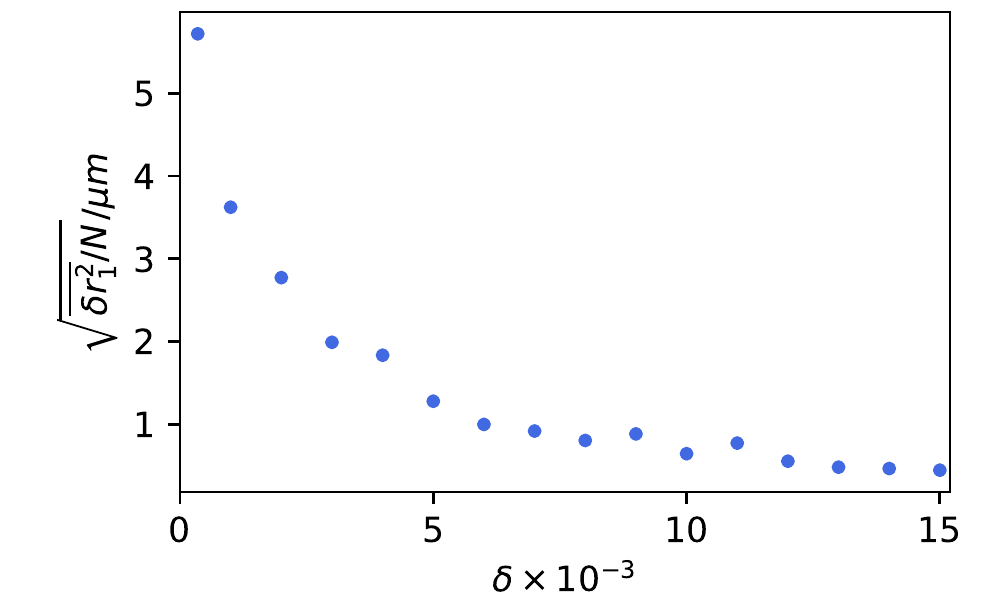}
 \caption{ \label{fig:sensitivity} Relation between the root mean square displacement $(\overline{\delta r^2_1}/N)^{1/2}$ of the rocking mode and the relative rotating wall strength $\delta$. The smallest value considered here is $\delta=3.5\times 10^{-4}$. Other parameters are the same as discussed in Sec. \ref{subsec:benchmarking}.}
\end{figure}

We plot $(\overline{\delta r^2_1}/N)^{1/2}$  versus rotating wall strength in Fig.~\ref{fig:sensitivity}. The behavior seen in Fig.~\ref{fig:sensitivity} can be explained qualitatively as follows. A strong rotating wall causes a difference between the trapping potential in the $x$ and $y$ directions in the rotating frame
\begin{equation}
\begin{aligned}
\psi_x=&\frac{1}{2}(\beta+\delta)m\omega_z^2 x^2,\\
\psi_y=&\frac{1}{2}(\beta-\delta)m\omega_z^2 y^2. 
\end{aligned}
\label{equ:asymmetry}
\end{equation}
For $\delta=0$, the trapping potential is azimuthally symmetric, resulting in a circular crystal with a zero-frequency rocking mode. With increasing $\delta$, the asymmetry in the trapping potential leads to an elliptic crystal that is squeezed along the axis corresponding to the stronger trapping potential (in this case, the $x$-axis in the rotating frame). The breaking of the azimuthal symmetry is accompanied by the rocking mode acquiring a non-zero frequency that increases with $\delta$. Correspondingly, the mean squared displacement $\overline{\delta r^2_1}$ associated with this mode decreases resulting in improved crystal stability.

\begin{figure}[h]
\includegraphics[width=\columnwidth]{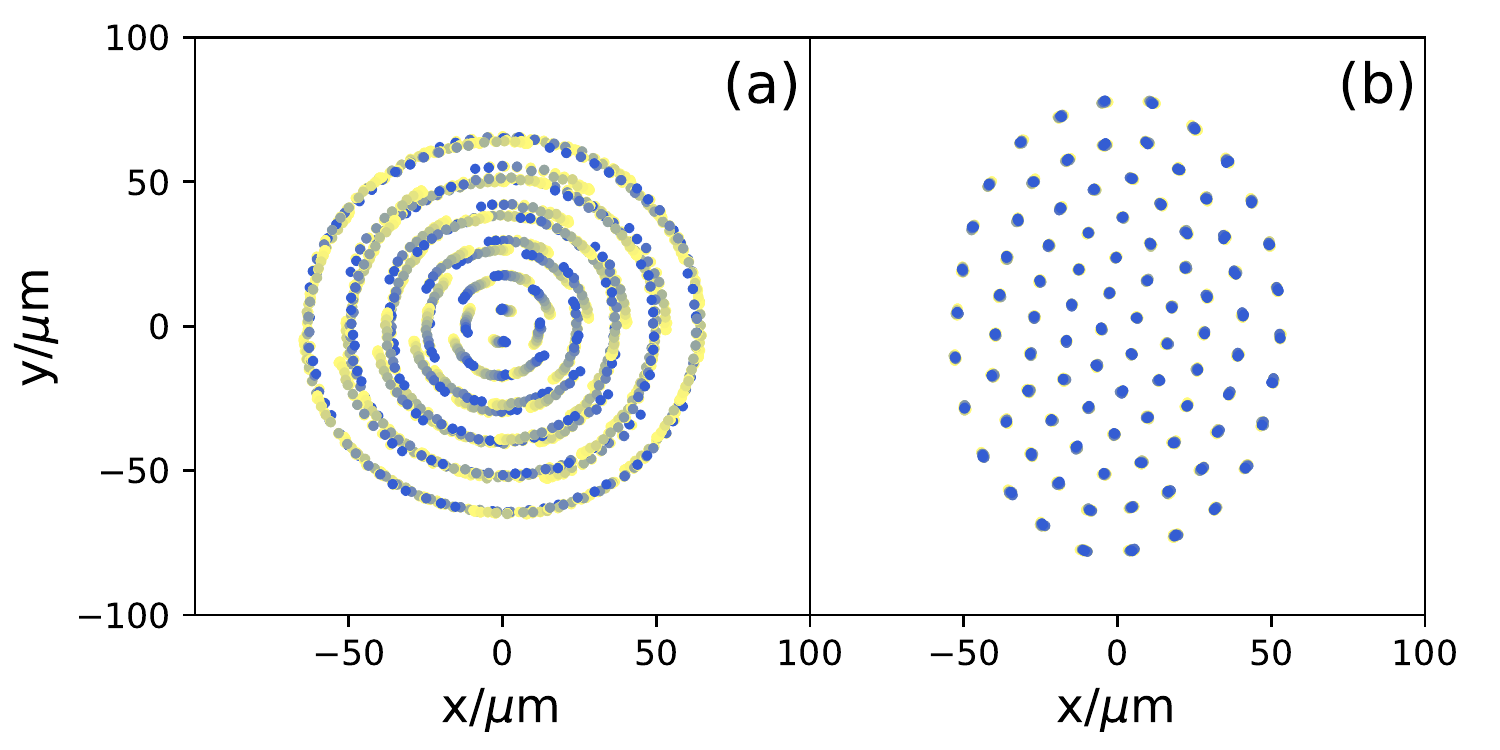}
 \caption{ \label{fig:traj} Time trace of ion trajectories in the rotating frame for crystals with $N=91$ ions. Recording duration is $\SI{1}{\milli\second}$ after the crystals have fully thermalized. We use a color gradient from yellow to blue to represent the chronological order. The relative rotating wall strengths are (a) $\delta=3.5\times 10^{-4}$ and (b) $\delta=0.0126$. Yellow dots in (b) are almost covered due to small displacements. Other parameters are the same as discussed in Sec. \ref{subsec:benchmarking}.}
\end{figure}

For illustration, we show the time trace of two crystal configurations with normalized wall strength of $\delta=3.5\times 10^{-4}$ and $\delta=0.0126$ in Fig.~\ref{fig:traj}. We first generate two equilibrium crystals with the respective rotating wall strengths and initialize their $\mathbf{E \times B}$ branches with $T_p=\SI{1}{\milli\kelvin}$. We then track the trajectories of the ions in the rotating frame once in thermal equilibrium (after $\SI{50}{\milli\second}$). A stronger rotating wall leads to a more stable configuration with well localized ions. In Fig.~\ref{fig:traj}, early times are represented by yellow dots and later times represented by blue dots. In Sec. \ref{sec:simulation}, we set  $\delta=0.0126$.

\begin{figure*}[htbp]
\includegraphics[width=17cm]{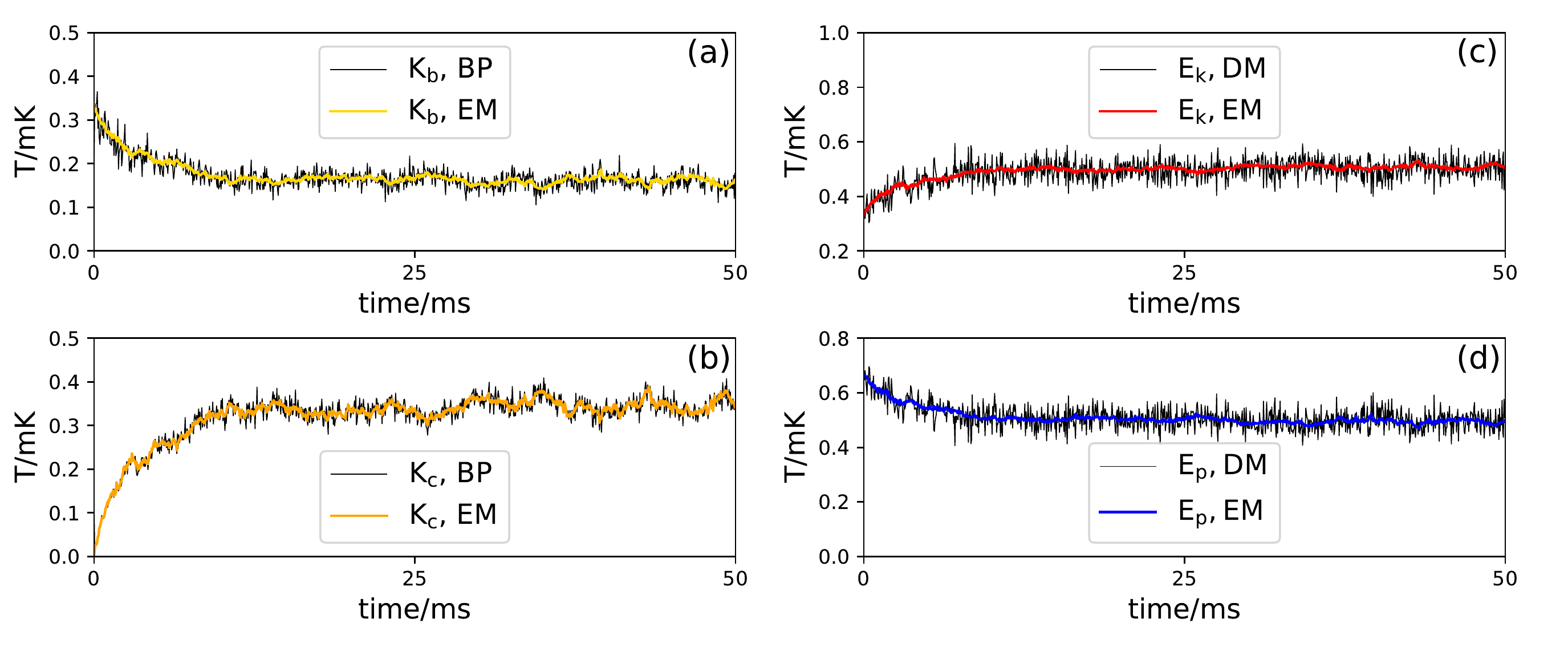}
\caption{\label{fig:validation}Eigenmode (EM), band-pass filter (BP), and direct (DM)  energy measurements during a $\SI{50}{\milli\second}$ evolution process after the $\mathbf{E \times B}$ branch is initialized with a homogeneous temperature of $\SI{1}{\milli\kelvin}$: (a) kinetic energy $K_b$ in $\mathbf{E \times B}$ branch; (b) kinetic energy $K_c$ in cyclotron branch; (c) total kinetic energy $E_k$ in the planar direction; (d) total potential energy $E_p$ in the planar direction. Other relevant parameters are reported in Sec.~\ref{subsec:benchmarking}.}
\end{figure*}

In passing, we note that besides ensuring the validity of the eigenmode method, crystals produced with a strong rotating wall may also offer several experimental advantages. The improved localization of the ions may be beneficial for implementing schemes for single-site addressing. The strong wall may also improve Doppler cooling of the planar modes, since torque from the cooling laser~\cite{torrisi2016PRA} can be more effectively counterbalanced, thereby ensuring that the crystal does not slip during the cooling process.

\section{\label{sec:simulation}Simulation of Planar Modes Coupling}

In this section, we perform molecular dynamics type simulations to study the coupling in the planar direction. During the thermal equilibration process, we validate the eigenmode measurement method by comparing the energy measurement results with the band-pass filter method. We then investigate the cyclotron-$\mathbf{E \times B}$ coupling as we vary $R$, the planar thermal temperature and the number of ions. Finally, we study the coupling within the $\mathbf{E \times B}$ branch and the cyclotron branch when the cyclotron-$\mathbf{E \times B}$ coupling is prohibited by large $R$.
\subsection{\label{subsec:benchmarking} Equilibration of the two branches}

Here, we present the thermal equilibration process using both energy measurement methods presented in Sec.~III. We generate a crystal of $N=91$ ions with charge $q=e$ in a Penning trap with parameters $R=5$, $\beta=0.05$, $\omega_r=2\pi\times\SI{400}{\kilo\hertz}$, $\delta=0.0126$, and $B=4.4588$ T. Accordingly, $\omega_z=2\pi\times\SI{0.704}{\mega\hertz}$, $\Omega=2\pi\times\SI{1.082}{\mega\hertz}$, and $m=63.3$ u. We generate an initial state far from the thermal equilibrium by initializing modes in only one of the two branches with non-zero thermal energy. Details of the initialization are discussed in Appendix \ref{app:initialization}. We initialize the 91 modes in the $\mathbf{E \times B}$ branch with a homogeneous temperature $ T_n=\SI{1}{\milli\kelvin}$ ($n\in b$), yielding $T_p=\SI{0.5}{\milli\kelvin}$. We then let the system evolve for $\SI{50}{\milli\second}$.

\begin{figure}[H]
\includegraphics[width=\columnwidth]{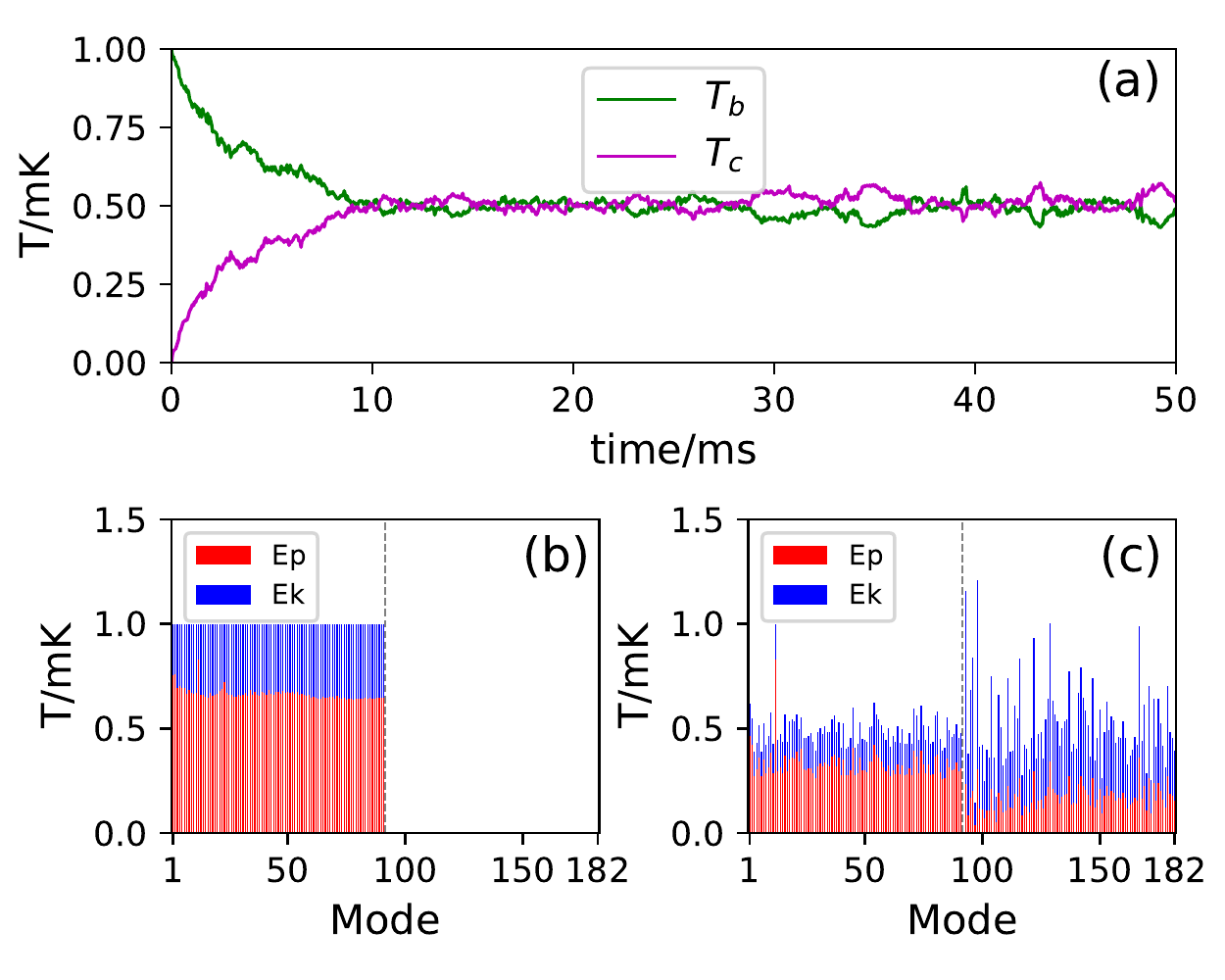}
 \caption{Thermal equilibration process in the planar direction with $R=5$: (a) Energies in the two branches ($T_b$ in green and $T_c$ in purple) during 50 ms evolution. (b) Energy distribution of  $2N=182$ modes at $t = 0$. (c) Energy distribution of $2N=182$ modes at $t=\SI{50}{\milli\second}$ averaged over 10 realizations with random-phase initial conditions (see Appendix~\ref{app:initialization}). The bars in (b) and (c) represent the kinetic (blue) and potential (red) energies of single modes in units of millikelvin. We note that the blue bars are stacked on top of the red bars and their total height gives the total energy of a mode. The 91 modes on the left and right sides of the gray dotted line belong to the $\mathbf{E \times B}$ and cyclotron branches, respectively. Other relevant parameters are reported in Sec.~\ref{subsec:benchmarking}.}
 \label{fig:coupling_example}
\end{figure}

\begin{figure*}
\includegraphics[width=18cm]{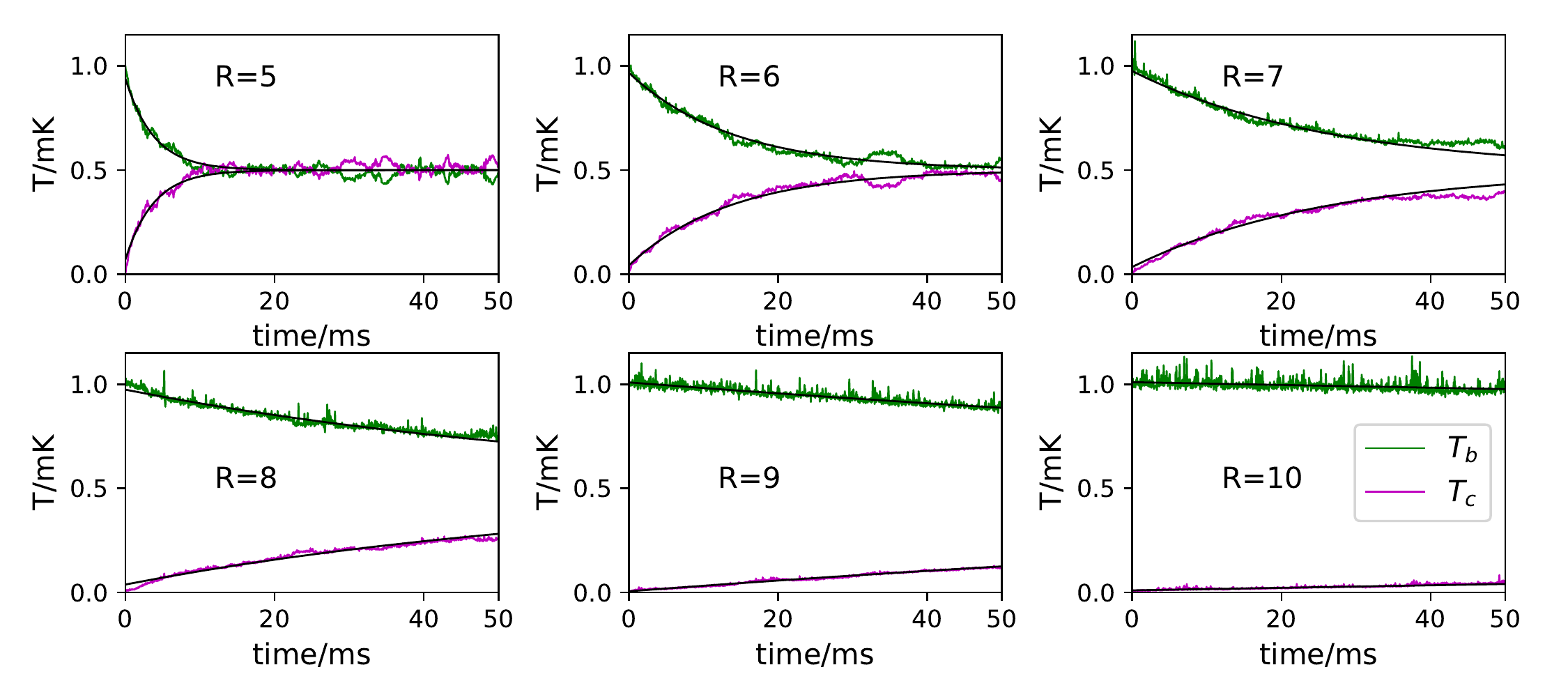}
 \caption{Time history of total energy in the two branches ($T_b$ in green and $T_c$ in purple lines) during $\SI{50}{\milli\second}$ evolution for values of $R$ varying from 5 to 10. Black lines are the exponential fitting function to determine the cyclotron-$\mathbf{E \times B}$ coupling rate $\alpha$. The initial temperature $T_b(0)=\SI{1}{\milli\kelvin}$.}
 \label{fig:R}
\end{figure*}

In Fig.~\ref{fig:validation}(a) and (b), we compare the kinetic energies in the two branches based on Eqs. (\ref{equ:Ek_EIG}) and (\ref{equ:Ek_FFT}).  From the frequency ranges in Fig. \ref{fig:Eval}, we use a filter with $l_0\Delta\omega=\SI{320}{\kilo\hertz}$ for the band-pass filter method. In Fig. \ref{fig:validation}(c) and (d), we compare the total kinetic and potential energies in the planar direction based on the eigenmodes method and a  direct measurement using the position and velocity coordinates of the ions. For the latter, we utilize $E_k=\sum m|\mathbf{v}_i|^2/2$ and Eq. (\ref{equ:HEP}) to directly measure the total kinetic and potential energies in the planar direction. The good agreement observed in Fig. \ref{fig:validation} demonstrates that the eigenmode method is valid at the low planar temperatures used in this paper.

Using the eigenmode method, we now plot the behavior of the total energies in the two branches in Fig.  \ref{fig:coupling_example}(a). We observe that the energies of the two branches approach $T_p$, which indicates an equipartition between the two branches. The dependence of the equipartition rate between the two branches on the parameters discussed in Sec. \ref{sec:coupling} is studied in the next section. To present details of the equipartition process, we compare the energy distribution in $2N=182$ modes at $t=0$ and 50 ms, as shown in Fig. \ref{fig:coupling_example}(b) and (c). The total energy for each $\mathbf{E \times B}$ mode is initialized at $\SI{1}{\milli\kelvin}$.  At later times, e.g. at $\SI{50}{\milli\second}$ as shown in Fig.~\ref{fig:coupling_example}(c), the system approaches equipartition.

\subsection{\label{subsec:coupling_rate}Dependencies of the cyclotron-$\mathbf{E \times B}$ coupling}

We now proceed to study the cyclotron-$\mathbf{E \times B}$ equilibration rate dependence on $R$, the initial temperature and the number of ions. We average every measurement over 10 realizations with random-phase initial conditions (Appendix~\ref{app:initialization}).

We measure the equilibration rate by fitting the time-dependent behavior of the temperatures in the two branches, $T_b=E_b/Nk_B$ and $T_c=E_c/Nk_B$, to the following exponential functions
\begin{equation}
\begin{aligned}
T_b(t) &= T_p\left[1+e^{-\alpha t}\right],\\
T_c(t) &= T_p\left[1-e^{-\alpha t}\right],
\end{aligned}
\label{equ:exponential}
\end{equation}
where we define $\alpha$ as the cyclotron-$\mathbf{E \times B}$ equilibration rate.  We will use this definition of $\alpha$ in what follows when we investigate the dependence of the equilibration rate on various parameters. To allow for a well-defined frequency gap between the two branches, we only investigate cases with $R\geq 5$. The following parameters are held constant: $\beta=0.05$, $\omega_r=2\pi\times\SI{400}{\kilo\hertz}$, $\delta=0.0126$, and $B=4.4588$ T.

\begin{figure}
\includegraphics[width=8cm]{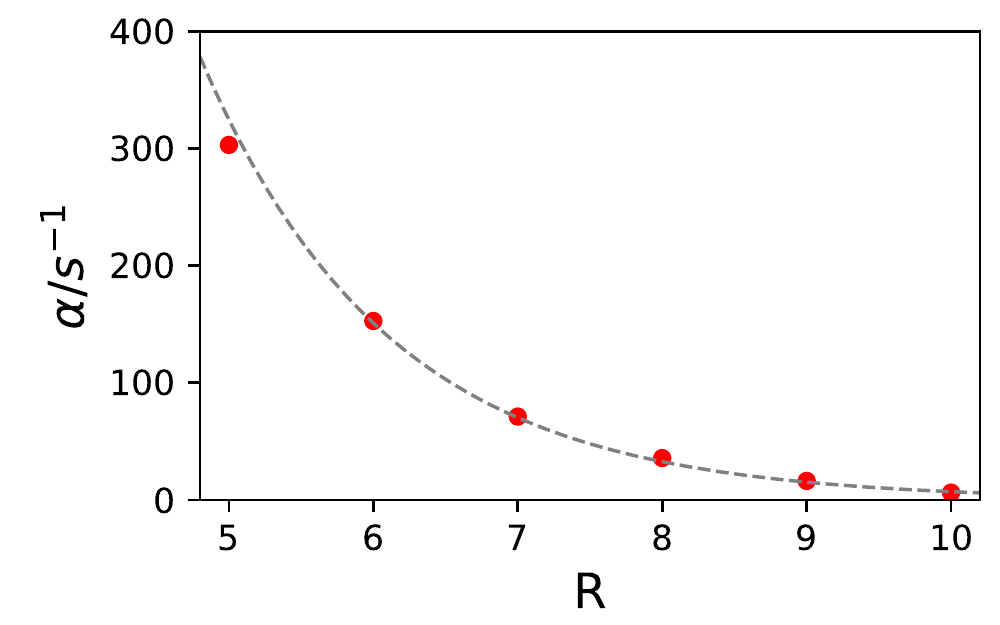}
 \caption{Measured coupling rate $\alpha$, obtained from Fig. \ref{fig:R}, as a function of $R$. The gray line is an exponential fit to the measurement results. Other parameters are the same as in Figs. \ref{fig:coupling_example} and \ref{fig:R}.}
 \label{fig:R_alpha}
\end{figure}

With $N=91$ we first vary $R$ from 5 to 10 with the 91 modes in the $\mathbf{E \times B}$ branch initialized with a homogeneous temperature $T_n\equiv \SI{1}{\milli\kelvin}$ ($n\in b$). The time histories of the energies in the two branches are shown in Fig.~\ref{fig:R}. We observe that, with increasing $R$, the time to equipartition increases. The black lines are exponential fits based on Eq. (\ref{equ:exponential}) that determine the equilibration rate $\alpha$. In Fig.~\ref{fig:R_alpha}, we display the relation between the fitted $\alpha$ and $R$. We find that $\alpha$ is exponentially suppressed with increasing $R$ with a  fitted exponential function (gray dashed line) dependence of $\alpha= \exp(-0.765R+9.608)\;\SI{}{\second}^{-1}$. This exponential scaling, showing suppression of the coupling rate with increasing ratio of frequencies is similar to what is seen in Ref. \cite{dan1993}. Moreover, the relevant parameters in current NIST experiments are $\omega_z=2\pi\times 1.585$ MHz, $\omega_r=2\pi\times 180$ kHz, $B=4.4588$ T, and $m({\rm Be}^+)=9.01$ u, resulting in $R=735$. For $N=91$ and assuming $T_p=\SI{0.5}{\milli\kelvin}$, we have $\alpha\sim 10^{-242} \;\SI{}{\second}^{-1} \sim 0 \;\SI{}{\second}^{-1}$. Such a small prediction for $\alpha$ suggests extremely weak coupling under current operating conditions of the NIST Penning trap. Any coupling will probably be due to other mechanisms such as mode interactions with error fields in the trap potential, which is not accounted for in our current model.

\begin{figure}
\includegraphics[width=9cm]{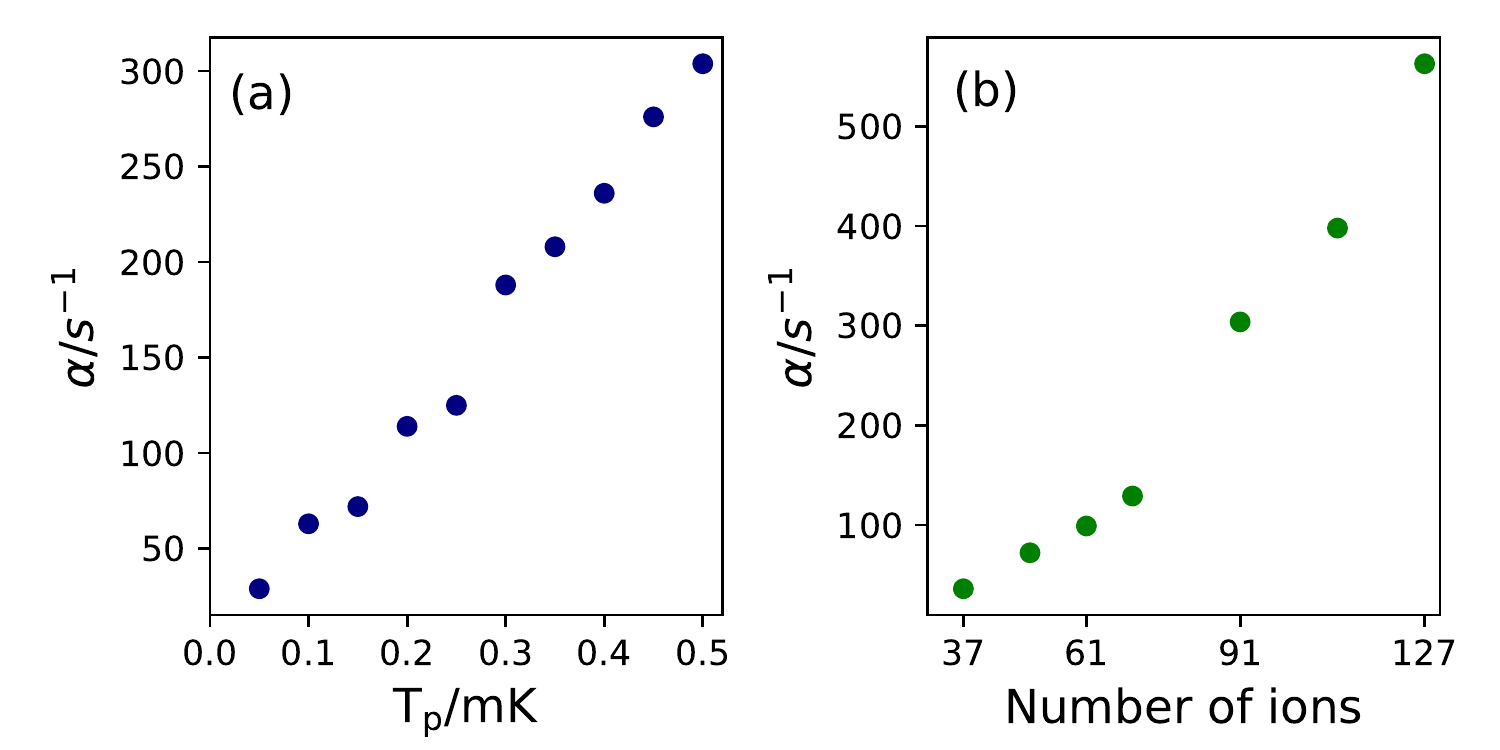}
 \caption{Measured coupling rate $\alpha$ as a function of (a) planar temperature (here the ion number $N=91$ is held fixed) and (b) the number of ions $N$ (here $T_p=\SI{0.5}{\milli\kelvin}$ is held fixed). Other parameters ($\delta, \beta, \omega_r, B$) are the same as in Sec. VI.A and Fig. \ref{fig:coupling_example}.}
 \label{fig:alpha_NT}
\end{figure}

We now fix $R=5$ and $N=91$ and study the dependence of $\alpha$  on $T_p$.  We perform similar simulations and exponential fitting as in Fig.~\ref{fig:R} to obtain the coupling rate for different $T_p$. As shown in Fig.~\ref{fig:alpha_NT}(a), as the planar temperature, $T_p$,  is varied from 0.05 to $\SI{0.5}{\milli\kelvin}$ we observe an approximate linear increase in $\alpha$.  Ions with higher temperature tend to have larger displacement, which increases the coupling rate.

Finally, we vary the number of ions from 37 to 127, while fixing $T_p\equiv \SI{0.5}{\milli\kelvin}$, to determine the dependence of $\alpha$ on $N$. Figure~\ref{fig:alpha_NT}(b) shows that $\alpha$ increases with the number of ions.

\subsection{\label{subsec:largeR}In-branch coupling for large $R$}

\begin{figure}[h]
\includegraphics[width=9cm]{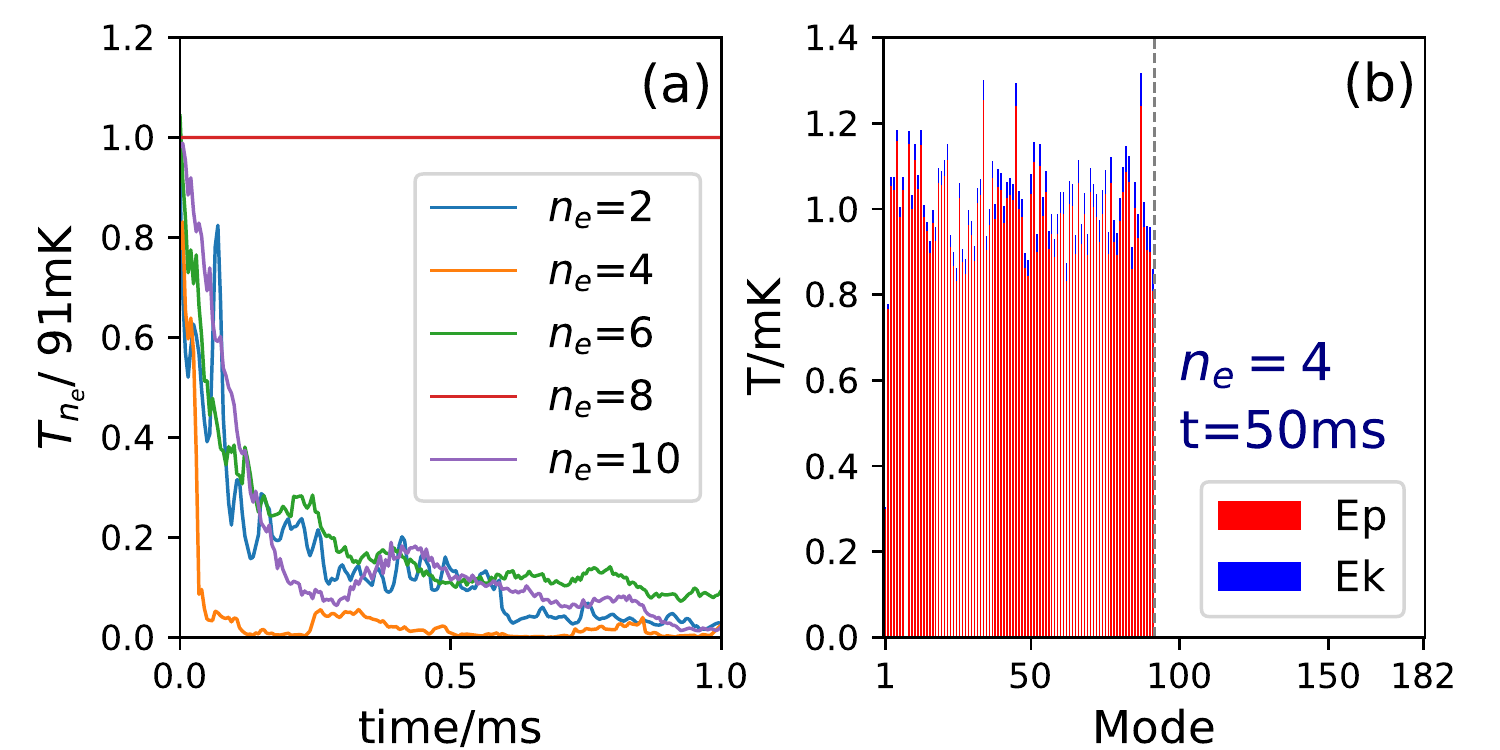}
 \caption{ \label{fig:exb_coupling}Coupling within the $\mathbf{E \times B}$ branch for $R=100$: (a) Normalized mode temperature $T_{n_e}/91$ of excited modes $n_e=2,4,6,8,10$ during independent evolutions. (b) Mode temperature distribution for the $n_e=4$ case at $t=\SI{50}{\milli\second}$. The 91 modes on the left and right sides of the gray vertical dotted line belong to the $\mathbf{E\times B}$ and cyclotron branches, respectively. Other parameters ($\delta, \beta, \omega_r, B$) are the same as in Sec. VI.A and Fig. \ref{fig:coupling_example}.}
\end{figure}

\begin{figure*}[htbp]
 \includegraphics[width=18cm]{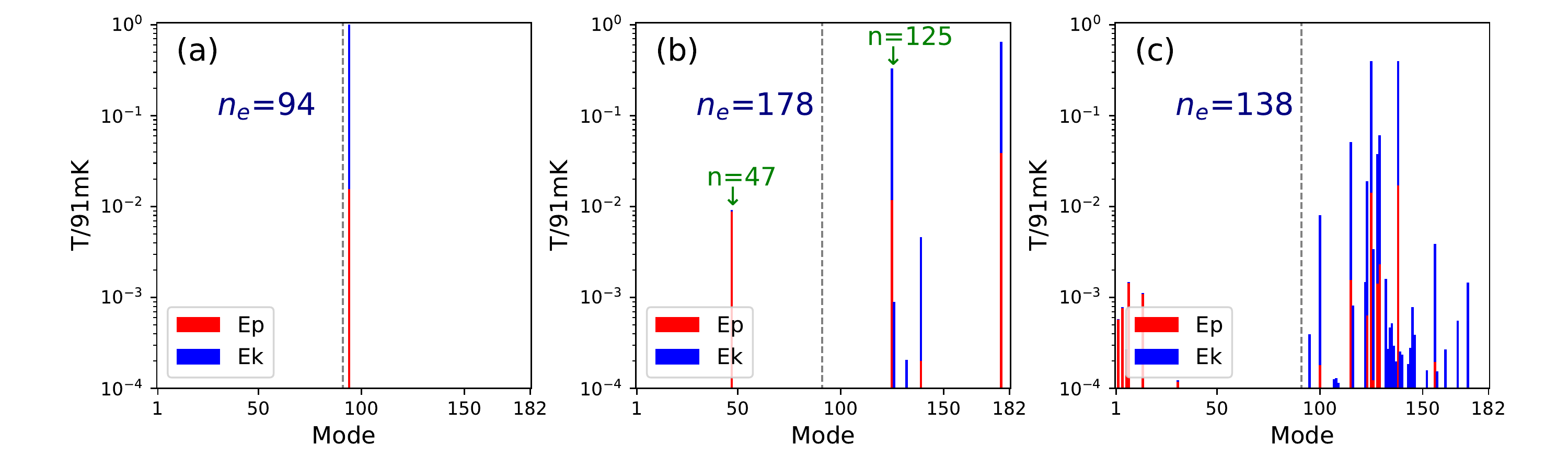}
  \caption{ \label{fig:cyc_coupling}Coupling within the cyclotron branch ($R=100$) for the cases where individual modes with (a) $n_e=94$, (b) $n_e=178$, or (c) $n_e=138$ are excited. Mode temperature distributions are measured at $t=\SI{50}{\milli\second}$. The 91 modes on the left and right sides of the gray dotted lines belong to the $\mathbf{E \times B}$ and cyclotron branches, respectively. In panel (b), the three modes with the highest temperatures are $n=$ 47, 125, and 178. Other parameters ($\delta, \beta, \omega_r, B$) are the same as in Sec. IV.A and Fig. \ref{fig:coupling_example}.}
\end{figure*}

\begin{figure*}[htbp]
 \includegraphics[width=18cm]{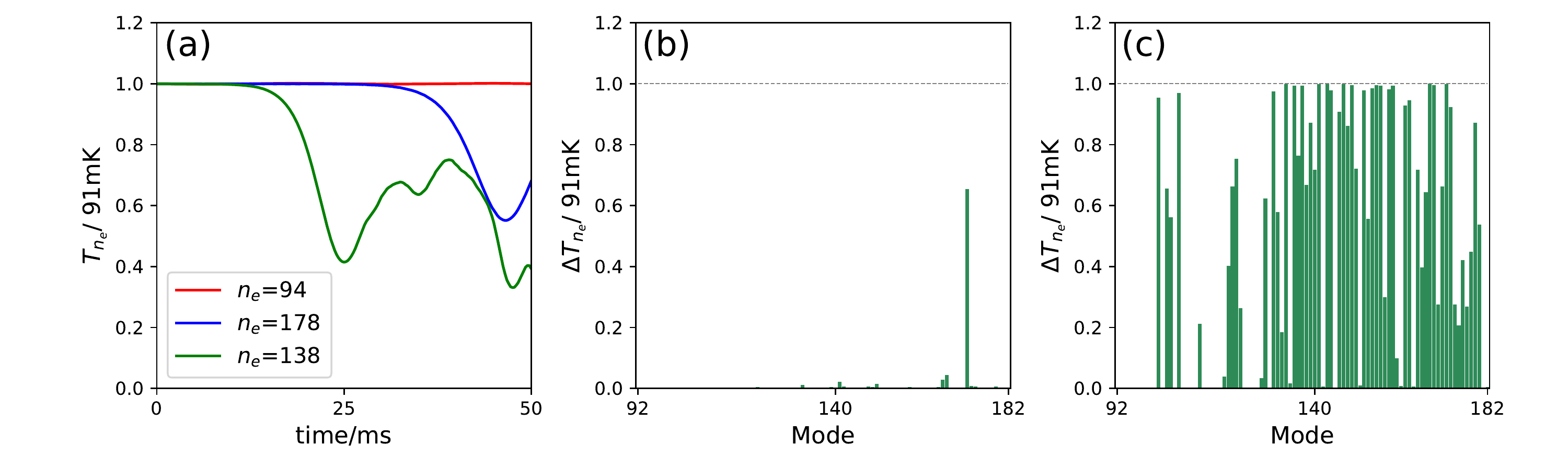}
  \caption{ \label{fig:fig13}  Coupling within the cyclotron branch for $R=100$. (a) Excited modes' temperatures for $n_e=$ 94, 178, and 138 during 50 ms evolution process. (b) Distribution of temperature range $\Delta T_{n_e}=\mathrm{max}(T_{n_e}(t))-\mathrm{min}(T_{n_e}(t))$ of $N=91$ initialized modes after evolution for $t=$ 10 ms. (c) $\Delta T_{n_e}$ distribution after $t=\SI{50}{\milli\second}$. Other parameters are the same as in Fig. \ref{fig:coupling_example}.}
\end{figure*}

Figure~\ref{fig:R_alpha} suggests an extremely weak cyclotron-$\mathbf{E \times B}$ coupling for a high value of $R$, which is consistent with the slow multi-phonon coupling process qualitatively discussed in Sec.~\ref{sec:coupling}. The large frequency gap and relatively small frequency ranges of the two branches make it impossible for a low temperature ($T_p<\SI{10}{\milli\kelvin}$) state to reach thermal equilibration on experimentally relevant timescales. In the absence of the equilibration of the two branches on the time scale of $\SI{50}{\milli\second}\ll 1/\alpha$, we can study the effect of in-branch coupling. In this section, we set $R=100$ while keeping other parameters ($\beta$, $\delta$, $\omega_r$, and $B$) the same as in Sec.~\ref{subsec:benchmarking}. To study the coupling within either branch, we only initialize one single mode $n_e$ for each initial state with mode temperature $T_{n_e} \equiv  N\times \SI{1}{\milli\kelvin}$  in order that the mean thermal temperature in the planar direction is still $T_p=\SI{0.5}{\milli\kelvin}$.

We first study the coupling among modes in the $\mathbf{E \times B}$ branch. During each evolution process we measure the temperature of the single initialized mode. In Fig.~\ref{fig:exb_coupling}(a) we present cases where individual modes with $n_e=2,4,6,8,10$ are initialized. Except for the case when the initialized mode is the center-of-mass mode ($n_e=8$, red line in Fig.~\ref{fig:exb_coupling}(a)), the temperature of the initialized mode decreases within 1 ms of evolution. The relative displacements of ions do not change under center-of-mass motion, making this mode immune to the Coulomb interaction. To investigate how the energy of the initialized mode is eventually distributed, we plot the energy distribution for the $n_e=4$ case at $t=\SI{50}{\milli\second}$ in Fig.~\ref{fig:exb_coupling}(b). We observe that the energy is approximately uniformly shared by the $\mathbf{E \times B}$ modes, but, as expected, the cyclotron modes are well isolated and no energy transfer happens between the two branches. The results in Fig.~\ref{fig:exb_coupling} indicate a strong coupling between modes within the $\mathbf{E \times B}$ branch.

We now proceed to study the coupling between modes in the cyclotron branch. In contrast to the $\mathbf{E \times B}$ branch, we find that the intrabranch coupling proceeds much more slowly. Figure~\ref{fig:cyc_coupling} shows some characteristic examples.  In Fig.~\ref{fig:cyc_coupling}, we plot the energy distribution at $t=\SI{50}{\milli\second}$ for the cases where individual modes with $n_e=94$, $178$, or $138$ were initialized with an initial temperature of $\SI{91}{\milli\kelvin}$. Some non-center-of-mass modes like the $n_e=94$ case shown in Fig.~\ref{fig:cyc_coupling}(a) are effectively decoupled from the other modes. On the other hand, Fig.~\ref{fig:cyc_coupling}(b) shows one of the simplest coupling mechanisms involving only three modes. In the $n_e=178$ case, the primary coupling involves two cyclotron modes ($n=178$, $\omega_{178}=2\pi\times 4.191$ MHz and $n=125$, $\omega_{125}=2\pi\times 4.076$ MHz) and one $\mathbf{E \times B}$ mode ($n=47$, $\omega_{47}=2\pi\times 0.115$ MHz) that satisfy a resonance condition, i.e. $\omega_{125}+\omega_{47}=\omega_{178}$. Although an $\mathbf{E \times B}$ mode is involved, this three-wave mixing process preserves the total phonon number in the cyclotron branch and hence cannot lead to thermal equilibration between the two branches~\cite{shijie1993}. We also present a multi-mode coupling in Fig.~\ref{fig:cyc_coupling}(c), in which several cyclotron and $\mathbf{E \times B}$ modes are excited.

To demonstrate that the coupling in the cyclotron branch is very slow, we measure and display in Fig.~\ref{fig:fig13}(a) the temperature of the single mode that was initialized in Figs.~\ref{fig:cyc_coupling}(a), (b), and (c). For $n_e=178$ and $138$ the mode temperature slowly changes during a 50 ms evolution time. From such plots, we can measure the temperature range $\Delta T_{n_e}=\mathrm{max}(T_{n_e}(t))-\mathrm{min}(T_{n_e}(t))$ sampled by the initialized mode $n_e$ during an evolution of duration $t$. In Fig.~\ref{fig:fig13}(b) and (c), we choose two time cutoffs ($t=\SI{10}{\milli\second}$ and $t=\SI{50}{\milli\second}$) and plot the distribution of $\Delta T_{n_e}$ when each cyclotron mode is separately initialized and allowed to evolve. For $t=10$ ms, most excited cyclotron modes are still isolated with their energy not transferred to other modes. As $t $ increases, more excited modes begin to exchange energy with other modes, but the intrabranch coupling is much slower compared to that within the $\mathbf{E \times B}$ branch.

In the case of the $\mathbf{E \times B}$ modes, a single initialized mode is typically observed to lose energy in an exponential manner. The other modes in the $\mathbf{E \times B}$ branch serve as an effective thermal reservoir leading to damping of the initialized mode on a timescale of a few tenths of a millisecond. However, in the case of the cyclotron branch, the time evolution of the energy in the initialized mode does not resemble exponential damping and instead shows signatures of revivals. In this case, the initialized cyclotron mode only couples to a few spectator modes on the timescale of the simulation, which is not sufficient to resemble an effective thermal reservoir of modes. The vast difference in the timescale of damping in the two branches may be attributed to the fact that the anharmonic terms in the Coulomb interaction scale with position fluctuations. For large values of $R$, position fluctuations are almost exclusively associated with the $\mathbf{E \times B}$ branch, and hence the in-branch equilibration is much faster here than in the cyclotron branch.

\section{\label{sec:discussion}Summary}

We have used an eigenmode analysis to study the thermal equilibration in the planar direction of a simulated two-dimensional ion crystal in a Penning trap with a rotating wall. We first solved for the eigenvectors and eigenvalues by linearizing the dynamics about the crystal equilibrium. We then validated the eigenmode analysis method by comparing with kinetic energies measured with a velocity filter technique and total energies calculated from a direct measurement of the ion positions and velocities. In the process, we discussed how a strong rotating wall helps reduce the amplitude of the rocking mode resulting in  a more stable crystal structure. To study the thermalization process in the planar direction, we initialized the modes in the low frequency $\mathbf{E \times B}$ branch with a specified temperature and performed first-principle simulations to measure the thermalization process. Finally, for large $R$ we studied the thermalization process within each branch by initializing the energy of a single mode and simulating the resulting equilibration process.

We investigated the dependence of the thermal equilibration rate between the cyclotron and $\mathbf{E \times B}$ branches on several trap and ion crystal parameters. We found that this equilibration rate is exponentially suppressed as a function of the ratio $R$ of the center-of-mass cyclotron to $\mathbf{E \times B}$ mode frequencies. The parameter $R$ provides a measure of the effective strength of the magnetic field on the dynamics of the in-plane motion \cite{Glinsky_1992}.  We also investigated the dependence of the cyclotron-$\mathbf{E \times B}$ equilibration rate on the planar temperature $ T_p$, and the number of ions, both of which exhibited an approximate linear dependence. In the simulations presented here, we fixed other aspects of the Penning trap, including the radial trapping strength $\beta$, rotating frequency $\omega_r$, rotating wall strength $\delta$, and the magnetic field $B$.

For large $R$ ($R = 100$), where the coupling between cyclotron and $\mathbf{E \times B}$ modes is very weak, we also investigated the internal coupling rate within the $\mathbf{E \times B}$ branch and within the cyclotron branch.  The $\mathbf{E \times B}$ branch was observed to rapidly equilibrate on a time scale of a few tenths of a millisecond.  The cyclotron branch equilibration time was more than two orders of magnitude longer and showed revivals instead of exponential damping.

Understanding planar equilibration and coupling between planar modes provides a starting point for understanding Doppler and sub-Doppler cooling in the planar direction. Doppler cooling of the $\mathbf{E \times B}$ modes is not well understood \cite{athreya2020}. Current NIST Penning trap experiments~\cite{britton2012engineered,bohnet2016quantum,sawyer2014spin} have $R\sim 735$, indicating that the $\mathbf{E \times B}$ branch is not cooled through a coupling to the cyclotron branch, which is efficiently cooled by Doppler laser cooling. The high frequency ratio $R$ also results in unequal energy distributions \cite{athreya2020}, in which energies in $\mathbf{E \times B}$ and cyclotron branches are predominantly potential and kinetic, respectively. An efficient cooling of the $\mathbf{E \times B}$ branch requires a cooling technique that can remove potential energy fluctuations associated with the ion positions. Axialization, which provides such a technique and has been carefully studied for single and small numbers of trapped ions~\cite{Thompson_2000, Thompson_2008}, may also work with many-ion crystals and will be the subject of future theoretical investigations. Finally, the long coherence time of the cyclotron modes motivates finding ways of employing these modes in quantum information processing. 

\section{Acknowledgement}


We thank Murray J. Holland, Wes Johnson, and Jennifer Lilieholm  for useful discussions and reading the manuscript. Work supported in part by U.S. Department of Energy (DOE) under award number DE-FG02-08ER54954 (C.T. and S.E.P.), the European Union's Horizon 2020 research and innovation programme under Grant Agreement No. 731473 (FWF QuantERA via QTFLAG I03769) (A.S.), the DOE Office of Science, National Quantum Information Science Research Centers, Quantum Systems Accelerator (QSA), the AFOSR Grant No. FA 99550-20-1-0019, and the DARPA ONISQ program (J.J.B.), and AFOSR Contract No. FA 9550–19-1–0999, DOE Grant No. DE-SC0018236, and NSF Grant No. PHY1805764 (D.H.E.D.).

\appendix

\section{\label{app:eigenmode}Linearization and eigenmode analysis}

Here we present the eigenmode analysis \cite{dan2020,athreya2020} of a two-dimensional crystal in a Penning trap. Because fluctuations are small for an ultracold ion crystal, one can linearize the ion motion about equilibrium positions $\mathbf{x}_{0i}=(x_{0i}, y_{0i},z_{0i})$ with small displacements $\mathbf{\delta x}_i=(\delta x_i,\delta y_i,\delta z_i)$. When the planar confinement is weak compared to that in the axial direction, the ion crystal is two dimensional \cite{wang2013}. The potential energy $\Psi$ is expanded at $\mathbf{x}_0$ using Taylor series to first order
\begin{equation}
\Psi=\Psi(\mathbf{x}_0)+\frac{1}{2} \sum_{ij} \mathbf{\delta x}_i \frac{\partial^2 \Psi}{\partial \mathbf{x}_i \partial \mathbf{x}_j} \mathbf{\delta x}_j.
\end{equation}
The out-of-plane (or axial) motion $\delta z$ in such a two-dimensional crystal is linearly decoupled from the planar motion $(\delta x,\delta y)$. The axial motion is described as a collection of $N$ simple harmonic normal modes \cite{athreya2020}. The normal modes in the planar direction, however, are not simple harmonic due to the velocity-dependent form of the Lorentz force. In this work, we solve for the normal modes in the planar direction. We begin by writing down the linearized equations for  $\mathbf{x}_{\perp}=(\delta x_1,...,\delta x_N,\delta y_1,...,\delta y_N)$ and $\mathbf{v}_{\perp}$=$d\mathbf{x}_{\perp}/dt$ as
\begin{equation}
\begin{aligned}
\frac{d\mathbf{v}_{\perp}}{dt}&=-\frac{\mathbb{K}_{\perp}}{m}\mathbf{x}_{\perp}+\mathbb{L}\mathbf{v}_{\perp}.
\label{equ:equ_planar}
\end{aligned}
\end{equation}
Here, 
\begin{equation}
\mathbb{K}_\perp=\frac{\partial^2 \Psi}{\partial \mathbf{x}^2} |_{\mathbf{x}=\mathbf{x}_0}
\end{equation}
is a real symmetric matrix \cite{wang2013} and $\mathbb{L}$ is the antisymmetric Lorentz force matrix ($2N\times 2N$) given by
\begin{equation}
\begin{aligned}
\mathbb{L}=(\Omega-2\omega_r)
\begin{bmatrix}
\mathbf{0}_N&-\mathbb{I}_N \\ 
\mathbb{I}_N & \mathbf{0}_N
\end{bmatrix}.
\end{aligned}
\end{equation}
We introduce the composite phase vector $\mathbf{u}_\perp=(\mathbf{x}_\perp,\mathbf{v}_\perp)^T$ and rewrite Eq. (\ref{equ:equ_planar}) as
\begin{equation}
\begin{aligned}
\frac{d\mathbf{u}_{\perp}}{dt}=\mathbb{D_{\perp}}\mathbf{u}_{\perp},
\end{aligned}
\label{equ:matrix}
\end{equation}
where  $\mathbb{D_{\perp}}$ is a composite matrix ($4N\times 4N$)
\begin{equation}
\begin{aligned}
\mathbb{D_{\perp}}=
\begin{bmatrix}
 \mathbf{0}_{2N}&\mathbb{I}_{2N} \\ 
-\frac{\mathbb{K}_{\perp}}{m} & \mathbb{L}
\end{bmatrix}.
\end{aligned}
\label{equ:matrix_form}
\end{equation}
We also combine the linearized potential energy and kinetic energy to obtain the total thermal fluctuation energy in the planar direction
\begin{equation}
\begin{aligned}
E=\frac{1}{2}\mathbf{u}_\perp^T \mathbb{E}\mathbf{u}_\perp=\frac{1}{2}\mathbf{r}_{\perp}^T \mathbb{K}_{\perp}\mathbf{r}_{\perp}+\frac{m}{2}\mathbf{v}_{\perp}^T \mathbf{v}_{\perp},
\end{aligned}
\label{equ:E_planar}
\end{equation}
where  
\begin{equation}
\mathbb{E}=\mathbf{diag}\{\mathbb{K}_\perp,m\mathbb{I}_{2N}\}
\label{equ:E_matrix}
\end{equation}
is the energy matrix in the planar direction.

Next, we solve Eq. (\ref{equ:matrix}) as an eigenvalue problem. We apply ansatz $\mathbf{u}_{\perp} =\mathbf{u}_{\omega} e^{-i\omega t}$ to transform Eq. (\ref{equ:matrix}) into $-i\omega \mathbf{u}_{\omega}=\mathbb{D_{\perp}}\mathbf{u}_{\omega}$. We then obtain $4N$ eigenvalues $\omega_n$ by solving the determinant equation $\mathrm{det}\|\mathbb{D}_\perp+i\omega_n\mathbb{I}_{4N}\|=0$. The elements of $\mathbb{D_{\perp}}$ and $\omega_n$ are all real \cite{dan2020} which results in pairs of complex conjugate eigenvectors, $\mathbf{u}_n$ and $\mathbf{u}_n^*$, associated with eigenvalues $\omega_n$ and $-\omega_n$, respectively. Therefore, there are $2N$ positive and distinct eigenvalues $\omega_n$ that represent the frequencies of $2N$ normal modes.

The eigenvectors are $\mathbb{E}$-orthogonal according to Ref. \cite{athreya2020} and \cite{dan2020}, which allows us to normalize the eigenvectors by means of
\begin{equation}
\begin{aligned}
\mathbf{u}_n\rightarrow \frac{\mathbf{u}_n}{\sqrt{\mathbf{u}_n^*\mathbb{E} \mathbf{u}_n}}.
\label{eqn:un_norm}
\end{aligned}
\end{equation}
The orthonormal eigenvectors satisfy $\mathbf{u}_m^*\mathbb{E} \mathbf{u}_n=\delta_{mn}$, where $\delta_{mn}$ is the Kronecker delta.

\section{\label{app:initialization} Initialization of ions}

To generate an initial state that is far from thermal equilibrium, we initialize one or several eigenmodes to create an inhomogeneous distribution of eigenmode energies. We perform the initialization in the lab frame, where a two-dimensional crystal in equilibrium is described by the coordinates $\mathbf{X}_0=(X_1,...,X_N,Y_1,..., Y_N)$, $Z_i=0$, and velocities $\mathbf{V}_0=\mathbf{\omega}_r\times \mathbf{X}_0$ corresponding to the collective rotation of all the ions. We also utilize the corresponding orthonormal eigenvectors $\{\mathbf{u}_n, n=1,...,2N\}$ that are determined in the rotating frame. As an example of the procedure, suppose we initialize one mode. We multiply the associated eigenvector with a random phase $e^{i\psi_n}$. We then take the real part as
\begin{equation}
\mathbf{U}_n=\mathrm{Re}\left[\mathrm{exp}(i\psi_r)\mathbf{u}_n  \right]
\label{equ:EM_one_mode}
\end{equation}
and decompose into the position and velocity parts as 
\begin{equation}
\mathbf{U}_\perp=(\mathbf{R}_\perp,\mathbf{V}_\perp)^T.
\label{eigen_one_mode}
\end{equation}
Next, we give each ion an extra displacement $\lambda \mathbf{R}_\perp$, where $\lambda$ is a normalization factor producing a desired thermal temperature $T_n$ for mode $n$ as
\begin{equation}
k_BT_n= \lambda^2 (\mathbf{R}_\perp^* \mathbb{K}_{\perp}\mathbf{R}_\perp + m\mathbf{V}_\perp^* \mathbf{V}_\perp).
\label{equ:EPK2}
\end{equation}
The resulting positions and velocities are $\mathbf{X}'=\mathbf{X}_0+\lambda \mathbf{R}_\perp$ and  $\mathbf{V}'=\mathbf{\omega}_r\times \mathbf{X}'+\lambda\mathbf{V}_\perp$. The initialization introduces a rotation $\lambda \mathbf{\omega}_r\times \mathbf{R} $ and produces an initial thermal distribution in the chosen mode. In addition, if we initialize multiple modes, we multiply each selected eigenvector with a random phase $e^{i\psi_n}$ and take the real part of the sum of the phase-multiplied eigenvectors
\begin{equation}
\begin{aligned}
\mathbf{U}_\perp=\mathrm{Re}\left[ \sum_{n\in \{b\}} e^{i\psi_n}\mathbf{u}_n  \right].
\end{aligned}
\label{equ:EM_initialization}
\end{equation}
We then decompose $\mathbf{U}_\perp$ in order to give ions extra displacement and velocities similar to Eqs. (\ref{eigen_one_mode}) and (\ref{equ:EPK2}) in the one mode case.

\bibliography{cite.bib}
\end{document}